\documentclass[letterpaper,aps,prl,twocolumn,floatfix,amsfonts,amssymb,notitlepage,superscriptaddress,longbibliography]{revtex4-2}

\pdfpageattr {/Group << /S /Transparency /I true /CS /DeviceRGB>>}

\usepackage{amsmath}
\usepackage{amsthm}
\usepackage{amssymb}
\usepackage{graphicx}
\usepackage{tabularx}
\usepackage{color}
\usepackage{xcolor}
\usepackage[normalem]{ulem}
\usepackage[colorlinks=true, linkcolor=blue, citecolor=blue, urlcolor=blue]{hyperref}

\newcommand{\be}{\begin{equation}}
\newcommand{\ee}{\end{equation}}
\newcommand{\bea}{\begin{eqnarray}}
\newcommand{\eea}{\end{eqnarray}}

\def\bs#1\es{\begin{split}#1\end{split}}	\def\bal#1\eal{\begin{align}#1\end{align}}

\begin{document}

\title{Metallic island array as synthetic quantum matter: \\fractionalized
entropy and thermal transport}

 \author{
Nitay Hurvitz}
\affiliation{Raymond and Beverly Sackler School of Physics and Astronomy, Tel Aviv University, Tel Aviv 69978, Israel} 
\author{Gleb Finkelstein}
\affiliation{Department of Physics, Duke University, Durham, NC 27701, USA}
\author{Eran Sela}
\affiliation{Raymond and Beverly Sackler School of Physics and Astronomy, Tel Aviv University, Tel Aviv 69978, Israel}


\begin{abstract}
\noindent 
The surprisingly rich physics of a single Coulomb-blockaded metallic island, when coupled to quantum Hall edge channels, is now well established -- giving rise to charge fractionalization and multi-channel quantum impurity behavior. Here, we show that qualitatively new physics emerges in arrays of such elements. We consider a 1D chain of $N$ metallic islands, focusing on thermodynamic signatures such as quantized entropy and anomalous thermal conductance. Universal and robust behavior emerges for energy scales smaller than the charging energy of the islands. In particular, we demonstrate that for the bulk filling factor of $\nu=1$, the islands could support a finite heat flow without temperature difference between them. Upon pinching the array with a quantum point contact, we predict an entropy change that scales with the number of islands as $\Delta S = \frac{1}{2}k_B \log (N+1)$, which can be measured using charge detection. This fractional entropy suggests the emergence of a novel type of excitations in the array.
\end{abstract}
\maketitle


In mesoscopic systems, ohmic contacts are typically three-dimensional (3D) metals used to interface low-dimensional quantum conductors—such as quantum Hall edge states with voltmeters or current sources. When the Ohmic contacts are themselves small, floating metallic islands on the micrometer scale, they become active components of the mesoscopic circuit. These islands exhibit significant charging energy effects while maintaining a continuous, metallic density of states.

When an island is connected to external reservoirs through weakly transmitting quantum point contacts (QPCs), only two charge states are relevant at low energies, effectively forming a charge pseudo-spin. Electron tunneling at the QPCs induces transitions between these states, analogous to spin flips in the Kondo problem~\cite{flensberg1993capacitance,PhysRevB.51.1743,furusaki1995coulomb}. The resulting charge Kondo effect has been realized using such metallic islands~\cite{iftikhar2015two,Iftikhar_2018,piquard2023observing}. This system exhibits a non-trivial ground state entropy characteristic of the multichannel Kondo effect—an entropy that has been theoretically linked to that of non-Abelian anyons~\cite{lopes2020anyons}.

Metallic islands also offer an efficient and controlled method for measurement of the heat conductivity of the quantum Hall edge states, which is quantized in units of quantum thermal conductance, $G_Q = \pi^2 k_B^2 T/3h$~\cite{jezouin2013quantum}. This technique was later extended to the non-Abelian $\nu = 5/2$ state, providing evidence for a half-integer thermal conductance~\cite{Banerjee_2017}.  Further careful low-temperature experiments have revealed the phenomenon of \textit{heat Coulomb blockade}~\cite{sivre2018heat}: the total heat flow emitted from an island connected to $n$ channels is found to be quantized at $n - 1$ units. This suppression arises because only the $n - 1$ \textit{neutral modes} contribute to heat transport, while the \textit{charge mode} is Coulomb blockaded and thus prevented from equilibrating with the island. Modeling these experimental findings builds on a seminal theoretical framework for understanding how quantum Hall edge states couple to or emerge from metallic islands~\cite{slobodeniuk2013equilibration} --- an approach that forms the foundation of our work.

A key motivation for this work comes from the theoretical analysis of Morel et al.~\cite{morel2022fractionalization}, who considered fractionalization of charge by a single metallic island connected to several edge channels. The case of a few islands has  been explored theoretically~\cite{PhysRevB.97.085403,karki2023z,Kiselev_2023,Nguyen_2024}, and in a pioneering experiment on two islands~\cite{pouse2023quantum}. Subsequent theoretical work suggests that this setup realizes an extended $Z_3$ quantum impurity, with a ground-state entropy $S= \frac{1}{2} k_B \log 3$ characteristic of parafermions~\cite{karki2023z}. Here we propose to scale these systems into arrays by coupling multiple floating islands as unit cells. 
In the regime where the charging energy exceeds the temperature, no fine tuning of the islands is required to generate a new type of synthetic quantum matter, whose properties are qualitatively different from those of its constituents. We will start with developing a bosonic scattering formalism, which we then apply to  thermodynamic signatures such as quantized entropy and anomalous thermal conductance. In particular, we demonstrate that for bulk filling factor of $\nu=1$, the chain of islands could support a finite heat flow \emph{without temperature difference between them}. We then introduce reflections and predict an unusual entropy scaling $\frac{1}{2}k_B \log (N+1)$ in a 1D chain of $N$ metallic islands upon partitioning it by a QPC.

\begin{figure}
\centering
\includegraphics[width=\columnwidth]{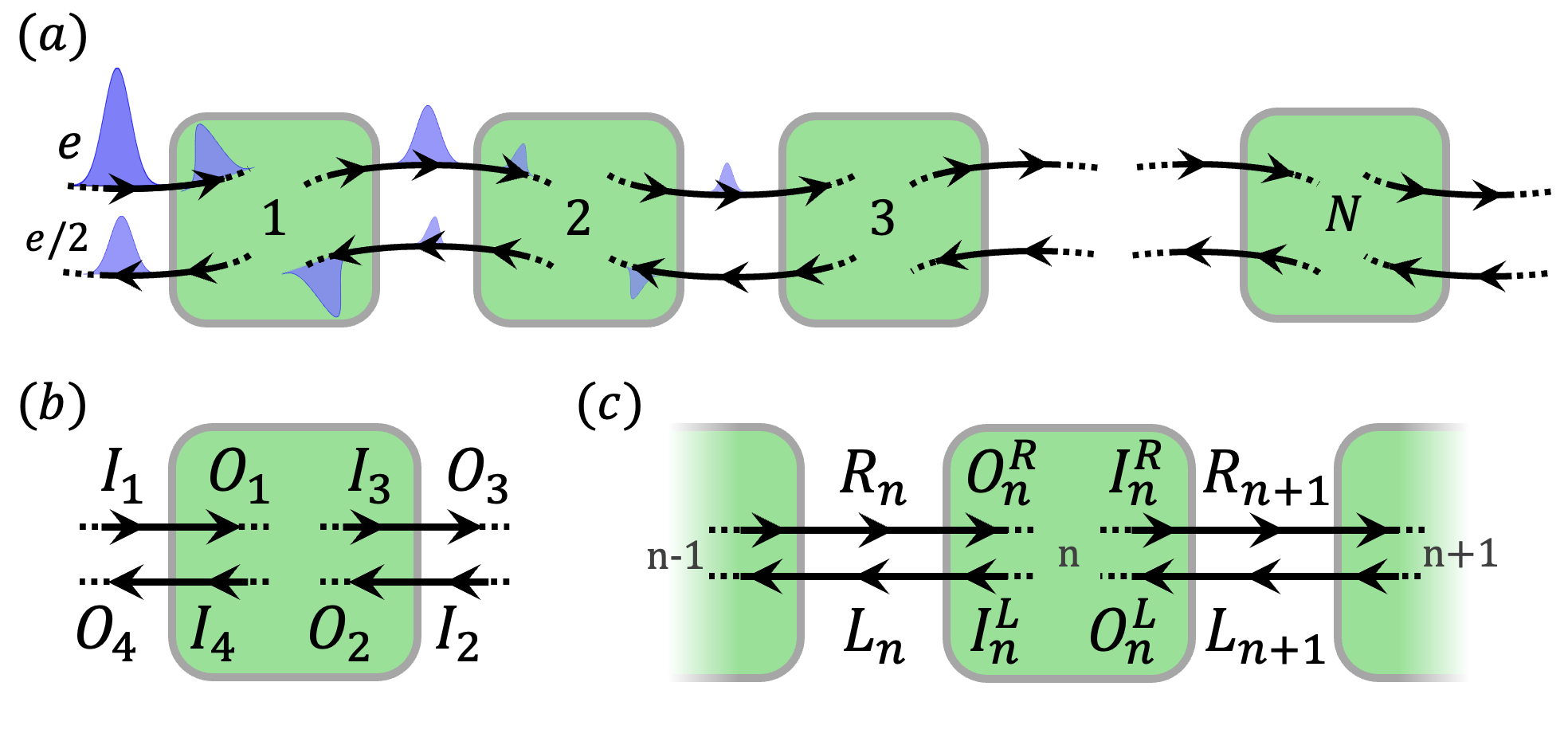}
\caption{(a) 1D array of metallic islands connected by quantum Hall edge
states. (b) Each island describes a scattering matrix Eq.~(\ref{eq:S}) connecting outgoing and incoming modes. Importantly, the channels extending into the island also act as either incoming $(I_{3,4})$ or outgoing $(O_{1,2})$ modes. (c) The properties of the array are obtained by combining  many such scattering matrices.
}
\label{fig:schematic}
\end{figure}

 \textbf{System:} We study a 1D array of floating metallic islands connected via a narrow strip of the quantum Hall material, Fig.~\ref{fig:schematic}.
Generalizing Refs.~\cite{slobodeniuk2013equilibration, morel2022fractionalization}, the Hamiltonian is $H = H_{\text{edge}} + H_{\text{charging}} + H_{\text{QPC}}$, where the edge states (here at $\nu=1$) connecting the islands are described by $H_{\text{edge}} = \frac{ v_F}{4\pi} \sum_{\text{edge } n=1}^{N+1} \int_{-\infty}^{\infty} dx \left[ (\partial_x \phi_{L,n})^2 + (\partial_x \phi_{R,n})^2 \right]$ with density operator $\rho_{R/L,n}=\pm \frac{1}{2\pi} \partial_x  \phi_{R,n}$ and free current operator $j_{R/L,n}=\pm v_F \rho_{R/L,n}=  \mp \frac{1}{2\pi} \partial_t  \phi_{R/L,n}
$. 
As illustrated in Fig.~\ref{fig:schematic}(a), each island $n$ is connected via two chiral edge modes $\phi_{L/R,n}$ to island $n-1$ ($n=0$ and $n=N+1$ refer to external contacts). The charging energy is given by $H_{\text{charging}} = \sum_{\text{island } n} E_c (Q_n - Q_{g,n})^2$. Here $Q_n$ is the number of electrons on the island $n$ and $Q_{g,n}$ is the offset controlled by the gate voltage.
The charging energy is assumed to be the dominant energy scale. Nonetheless, the charge on the islands is not quantized as long as the connecting channels have perfect transmission. Therefore, no fine-tuning of $Q_{g,n}$ is required, and these parameters will be set to zero unless specified otherwise.

In terms of the bosonic variables, the charge on island $n$ is given by $Q_n = \frac{1}{2\pi} \left[ \varphi_{n+1}(-a) - \varphi_{n}(a) \right]$, where $\varphi_n(x) = \phi_{R,n}(x) - \phi_{L,n}(x)$ and the points $x=-a(+a)$ are taken as the boundaries of island $n$ (or $n+1$) on the line connecting island $n$ and $n+1$.  We will take $a \to 0$ meaning that the level spacing in the links $\frac{\pi v_F}{2a}$ is large compared to the temperature. Finally, adding a QPC on link $n$ connecting islands $n-1$ and $n$ introduces a back-scattering term $H_{\text{QPC}} = \sum_{n=1}^{N+1} r_n \cos[\varphi_n(0)]$. We will start our consideration with a simple case of open QPCs, $r_n=0$. 

\textbf{Scattering formalism:} When all $r_n = 0$, the problem reduces to the scattering of bosons and can be solved at any frequency $\omega$ 
along the single-island methods of
Refs.~\cite{slobodeniuk2013equilibration,morel2022fractionalization}. For $\omega \gg E_c$, the charging energy becomes negligible and all edge modes are fully and independently transmitted into or out of the islands. In the opposite regime, $\omega \ll E_c$, the constrained total charge on the island imposes the condition $\varphi_{n+1}(0)=\varphi_{n}(0)$. We will be concerned with the second limit which displays novel effects of fractionalization.

Qualitatively, consider an incoming electron from the left as in Fig.~\ref{fig:schematic}(a). Due to the large charging energy $E_c$, the boundary conditions enforce equal fractionalization of the incoming charge into each of the island’s output channels, consisting of charge wave-packets of $e/2$. As we will see from  the scattering matrix in Eq.~(\ref{eq:S}) below, a $\pm e/2$ dipole also penetrates into the island. This process repeats through the array until a steady state is reached.

We first consider the scattering of the four fields from the boundary of a single island, Fig.~\ref{fig:schematic}(b). Following Ref.~\cite{oreg2014fractional}, we denote the currents with respect to the boundary, that is, denoting $I_i$ the current going \textbf{towards the boundary} and by $O_i$ the current outgoing \textbf{from the boundary}. Outside of the island there are 2 incoming currents ($I_1,I_2$) and 2 outgoing currents ($O_3,O_4$), there are also the 4 corresponding currents from inside the island: ($I_3,I_4)$ and  ($O_1,O_2$). Following Ref.~\cite{morel2022fractionalization} we define a rotated set of fields for each of the above pairs, corresponding to a charged mode ('+' superposition) and a neutral mode ('--' superposition). With the island's total charge being frozen for large $E_c$, the charged modes are fully reflected by the island's boundary, whether they go towards the island or from within it. In contrast, the neutral modes are fully transmitted, since they are decoupled from the island's total charge. From this follows the scattering matrix
\begin{equation}
\label{eq:S}
    S =  \frac{1}{2}\begin{pmatrix}
    1 & -1 & 1 & 1\\
    -1 & 1 & 1 & 1\\
    1 & 1 & 1 & -1\\
    1 & 1 & -1 & 1\\
    \end{pmatrix},
    ~~~~
    \begin{pmatrix}
    O_1 \\ O_2 \\ O_3 \\ O_4
    \end{pmatrix}
    = S \begin{pmatrix}
    I_1 \\ I_2 \\ I_3 \\ I_4
    \end{pmatrix},
\end{equation}
 relating the outgoing currents $O_i$ to the incoming currents $I_i$ for $\nu=1$, see \textit{Supplementary Material} for details.

Next we move to the 1D chain of islands, where the interaction of the currents with each island is described by the above $S$ matrix.
Each pair of neighboring islands $n-1$ and $n$ shares 2 currents, such that an outgoing current of one island is the incoming of the other. We thus relabel the currents to the left of the $n$-th island as $R_n$ and $L_n$ for the right- and left-moving currents, see Fig.~\ref{fig:schematic}(c).
The other pair of incoming and outgoing currents outside of island $n$ are $R_{n+1}$ and $L_{n+1}$, which are shared with its neighbor on the right.
The 4 currents inside the island retain $I$ and $O$ notation, with superscript $L,R$ indicating their direction: $O^L_n,O^R_n$ describe the currents moving away from the boundary to the left and the right, and $I^L_n,I^R_n$ describe the left- and right-going currents moving towards the boundary. The mapping from the previous notation is: $(O_1,O_2,O_3,O_4) \to (O^R_n, O^L_n ,R_{n+1} ,L_{n} )$ and similarly $(I_1,I_2,I_3,I_4) \to (R_n ,L_{n+1} ,I^R_n ,I^L_n)$.

The main implication of the scattering relation (\ref{eq:S}) is drawn from its two bottom equations, as they deal with the output currents going outside the island. 
Together with the conservation of the net current $\mathcal{I}$ through the array (i.e. $\forall n:~ \mathcal{I}= R_n-L_n$) we find the recurrence relations between currents moving in the same direction
\begin{equation}
\label{eq:recurrence}
        R_{n+1} = R_n + I^{(isl)}_n - \mathcal{I} ~~~;~~~ L_{n+1} = L_n + I^{(isl)}_n - \mathcal{I},
\end{equation}
with $I^{(isl)}_n \equiv I^R_n - I^L_n$ the net current from \textbf{within} the $n$-th island.
Both relations are solved by induction to find the currents in any of the channels in terms of the independent incoming currents only 
\begin{subequations}
\label{eq:sol_recurrence}
\begin{equation}
\begin{split}
    L_n = \frac{1}{N+1}
    \Bigg(
    &(n'-1)\bigg(R_1 + \sum_{i=1}^{n-1}I^{(isl)}_i\bigg)\\
    + 
    &n \bigg(L_{N+1} - \sum_{i=n}^{N}I^{(isl)}_i\bigg)
    \Bigg),
\end{split}
\end{equation}
\begin{equation}
\begin{split}
    R_n = \frac{1}{N+1}
    \Bigg(
    &n'\bigg(R_1 + \sum_{i=1}^{n-1}I^{(isl)}_i\bigg) \\
    + &(n-1) \bigg(L_{N+1} - \sum_{i=n}^{N}I^{(isl)}_i\bigg)
    \Bigg),
\end{split}
\end{equation}
\end{subequations}
with $n' = (N+1)-n+1$ being the index $n$ counted from the end of the array. 
See \textit{Supplementary Material} for a detailed solution. Importantly, the incoming fields have totally independent thermal and quantum fluctuations. Since there are $2(N+1)$ such currents, we can denote them by a vector
\be
\label{eq:incoming_vector}
\vec{I}=(  R_1,
        I^L_1,I^R_1,
        \dots ,
        I^L_N ,
        I^R_N ,
        L_{N+1}).
\ee

Since the solutions in Eq.~(\ref{eq:sol_recurrence}) are linear in the incoming currents, any linear combination of currents could then be expressed as $ \vec{d}\cdot\vec{I}$ with $\vec{d}$ being some vector of coefficients. For example, we can verify that the current though each link follows a very simple expression
\begin{equation}
\label{eq:L_R}
\begin{split}
   \mathcal{I}= R_n - L_n
    &= \frac{1}{N+1}(R_1 - \sum_{j=1}^N I^{(isl)}_j- L_{N+1})
    \\& = \frac{1}{N+1}
    \begin{pmatrix}
        1 & -1 & 1 & \cdots & -1 & 1 & -1
    \end{pmatrix}
    \cdot\vec{I},
\end{split}
\end{equation}
which does not depend on $n$, as expected.

As a first application of the formalism, let us discuss the process of charge transport. We consider the average of the  electric current $\mathcal{I} = R_n - L_n$. Using the fact that on average $\langle I_{n}^L \rangle=\langle I_{n}^R \rangle$, the average of Eq.~(\ref{eq:L_R}) gives $\langle \mathcal{I} \rangle = \frac{\langle R_1 - L_{N+1} \rangle}{N+1}$. According to Landauer formula, for the external reservoirs $\langle R_1 \rangle = \frac{e^2}{h}V_1$ and $\langle L_{N+1} \rangle = \frac{e^2}{h}V_2$. 
As a result, the DC conductance of the array is simply
\be
\label{eq:G}
G_N=\frac{1}{N+1}\frac{e^2}{h}.
\ee
While our primary interest lies in the nontrivial consequences of charge fractionalization, the DC electric conductance provides a useful reference point that remains entirely unaffected by the charging energy.

We note that the assumed absence of charge fluctuation in the islands led us to conclude that the current operator $\mathcal{I}$ is uniform along the chain. However a more careful treatment is required in the limit of large  $N$, where the multiple links should lead to the delayed propagation of the current fluctuations between the islands.

\begin{figure}
\centering
\includegraphics[width=\columnwidth]{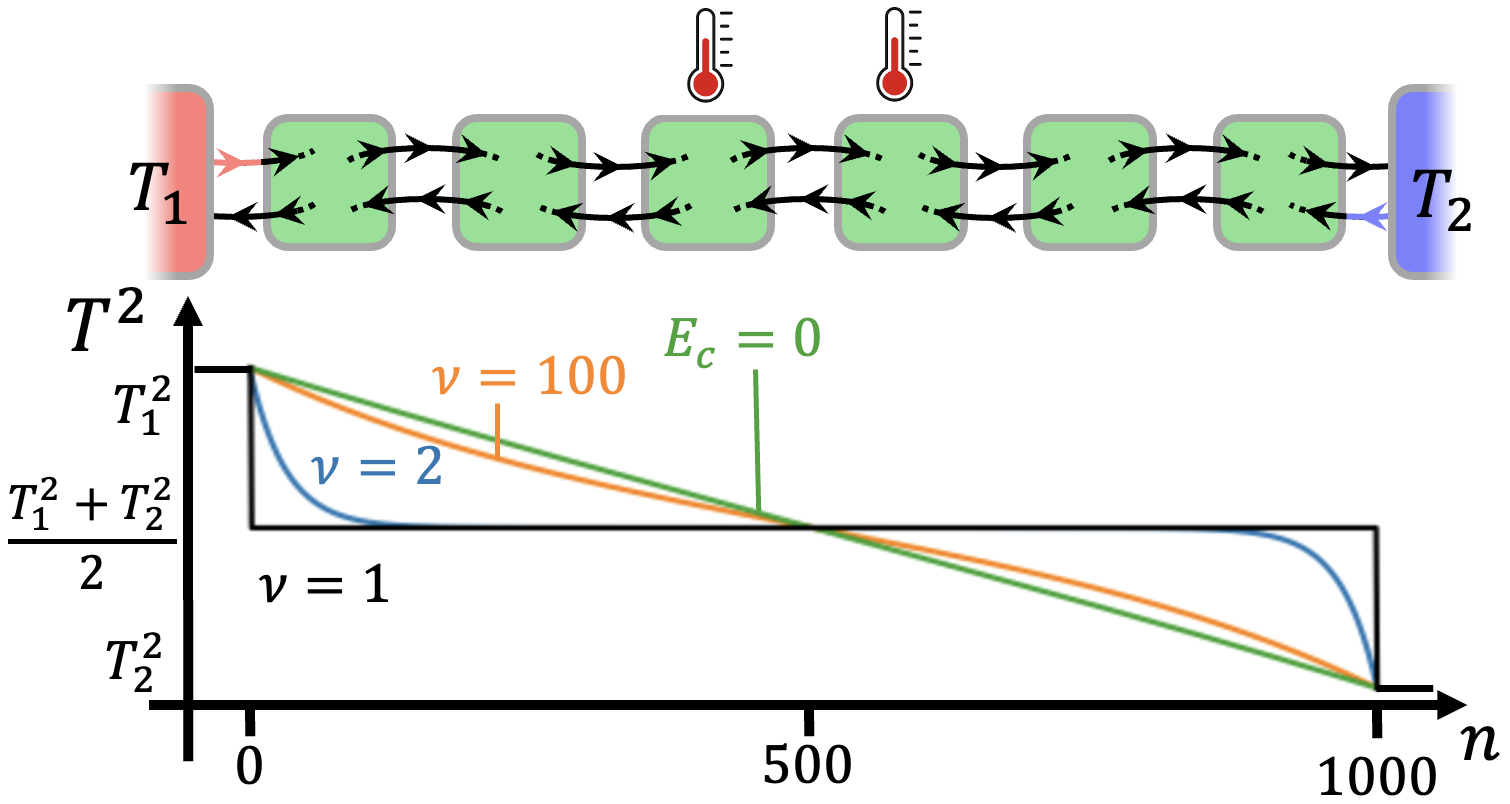}
\caption{Top: schematic of $N$ islands connected to microscopic reservoirs at temperatures $T_1$ and $T_2$. Bottom: temperature profile along the 1D array displaying constant temperature profile for $\nu=1$, which is a consequence of the heat Coulomb blockade. As the filling factor is increased, the temperature profile approaches that without the charging energy (see the $\nu=100$ curve), however for small $\nu$ it remains essentially flat in the middle of the array.}
\label{fig:heat_flow}
\end{figure}

\textbf{Heat flow and island temperature profile:} Consider connecting the array to a hot and cold reservoirs at its far ends as in Fig.~\ref{fig:heat_flow}. We are interested in characterizing the heat flow $J$ and the equilibrium temperature distribution along the array. The heat flow along a ballistic channel is proportional to the square of the current in that channel $J=\frac{\hbar \pi}{e^2} j^2$ ~\cite{slobodeniuk2013equilibration}. For an incoming mode, it is also proportional to the squared temperature of its source, $J=\frac{\pi^2}{6 \hbar}k_B^2 T^2 \equiv \kappa T^2$. Since incoming currents are independent ($\langle I_\mu I_\nu \rangle \propto\delta_{\mu\nu}$), if an outgoing or link current is written in terms of the incoming currents as $L_n = \sum_\mu d_{L_n,\mu} I_\mu$ and similarly for $R_n$, then the average heat current through that channel can be written as
\begin{equation}
\label{eq:heat_current_fluctutation}
\begin{split}
    \langle J_{L_n} \rangle 
    =\frac{\hbar \pi}{e^2} \langle L_n^2  \rangle 
    =\frac{\hbar \pi}{e^2} \sum_\mu 
    d_{L_j,\mu}^2 \langle I_\mu^2 \rangle 
    =\kappa \sum_\mu d_{L_j,\mu}^2 T_\mu^2,
\end{split}
\end{equation}
with 
$T_\mu$ the temperature of the source of the $\mu$-th incoming mode. We assume that the incoming currents $R_1$ and $L_{N+1}$ carry the temperature of the left and right external baths, and $I_n^R$ and $I_n^L$ carry the temperature $T_{\Omega_n}$ of the $n-th$ island.
To solve for the temperature profile $\{ T_{\Omega_n}\}$ we require that the sum of heat flowing from island $n-1$ to $n$ should be constant throughout the chain. 

In the simple case of a single island, there are 4 incoming currents, using the scattering relation Eq.~(\ref{eq:sol_recurrence}) %
we find the heat flow from the island to the left reservoir
\begin{equation}
\begin{split}
    L_1 & = \frac{1}{2}(R_1 - I^{(isl)}_1+L_2) \\ 
    &~~\rightarrow~~
    \langle J_{L_1} \rangle 
    = \frac{\kappa}{4}(T_1^2 + 2T_\Omega^2 + T_2^2).
\end{split}
\end{equation}
Note that the notation of $I^{(isl)}_n$, the net current from \textbf{within} the island, contains 2 independent currents $I^R_n, I^L_n$, which thus contribute $2\kappa T_{\Omega_n}^2$ to any heat current. $R_1$ is itself an incoming current, so the net heat flow in the left link is
\begin{equation}
    \langle J_1 \rangle 
    = \langle J_{R_1} \rangle  - \langle J_{L_1} \rangle 
    = \frac{\kappa}{4}(3T_1^2 - 2T_\Omega^2 - T_2^2).
\end{equation}
Similarly for the right link $
\langle J_2 \rangle 
= \frac{\kappa}{4}(T_1^2 + 2T_\Omega^2 - 3T_2^2)$. By equating the two heat currents, the island temperature is found to be $T^2_\Omega = \frac{T_1^2 + T_2^2}{2}$, which leads to a heat flow of $\langle J \rangle  = \frac{\kappa}{2} (T_1^2 - T_2^2)$.

We then proceed to the two island case with 6 independent currents. The channel currents are similarly calculated from Eq.~(\ref{eq:sol_recurrence}), e.g. $L_1 = \frac{1}{3}(2R_1 - I^{(isl)}_1 - I^{(isl)}_2 + L_3)$. Proceeding as in the one-island case to compute the heat currents along the individual links and demanding heat current conservation, we 
surprisingly obtain that the temperature of the two islands is the same as in a 1-island case:
\begin{equation}
T_{\Omega_1}^2 = T_{\Omega_2}^2 = \frac{T_1^2 + T_2^2}{2}.    
\end{equation}

Finally for $N$ islands, we use Eq.~(\ref{eq:sol_recurrence}) to find the heat currents $\langle J_{R_n}\rangle,\langle J_{L_n} \rangle$ for any $n$. Then the heat current $\langle J_n\rangle = \langle J_{R_n}\rangle-\langle J_{L_n} \rangle$ through any link $n$ follows immediately and is no more than a linear combination of squared temperatures,
\begin{equation}
\label{eq:heat_current}
\begin{split}
   \langle J_n \rangle = 
    \frac{\kappa}{(N+1)^2}
    \Bigg((2n'-1)(T_1^2 + 2\sum_{i=1}^{n-1} T^2_{\Omega_i})\\
     - (2n-1)(T_2^2 + 2\sum_{i=n}^N T^2_{\Omega_i})\Bigg),
\end{split}
\end{equation}
with $n' = (N+1)-n+1$ being the index $n$ counted from the end of the array.
Imposing equality through all heat currents by requiring $\langle J_{n+1} \rangle-\langle J_{n} \rangle =0$ we indeed get a constant temperature profile
\begin{equation}
\label{eq:T_island}
    T_{\Omega_n}^2 = \frac{T_1^2 + T_2^2}{2},
\end{equation}
just like the 1- and 2-island examples. Such temperature profile, plugged back into Eq.~(\ref{eq:heat_current}) results in a heat current of
\begin{equation}
    \langle J\rangle = \kappa\frac{T_1^2 -T_2^2}{N+1}.
\end{equation}

We stress that our result shows that the array acquires a constant temperature profile despite the presence of the heat flow, see Fig.~\ref{fig:heat_flow}. 
This result may be highly un-intuitive, and it signifies the effect of the heat-Coulomb blockade along the 1D array at $\nu=1$. The absence of temperature gradient within the array can be measured by attaching temperature sensors to the individual islands in the middle of the array. As the left reservoir is heated (schematic in Fig.~\ref{fig:heat_flow}), we expect the temperature of the islands to rise, reaching the same value given by Eq.~(\ref{eq:T_island}). The resulting ``four-probe'' measurement should therefore yield infinite heat conductivity.

The result above holds strictly for $\nu=1$. 
In order to generalize it  to higher integer filling factors ($\nu >1$) we need to replace each link current $L_n, R_n$ with $\nu$ different currents. Going along the same lines as before, these $\nu$ different modes could be rotated to form a single charged mode and $(\nu-1)$ neutral modes. While the charged mode obeys the same scattering relations, the neutral modes can only thermalize neighboring islands, carrying the heat of each island only to its adjacent neighbors, thus weakening the non-local effects. In order to adapt Eq.~(\ref{eq:heat_current}) for $\nu>1$ we have to add these $(\nu-1)$ heat currents flowing from the 2 islands connected by the link $n$. The resulting  heat current is
\begin{equation}
    \langle J_n^{(\nu>1)} \rangle = \langle J_n \rangle +(\nu-1)\cdot\kappa(T_{\Omega_{n-1}}^2 - T_{\Omega_n}^2),
\end{equation}
where $\langle J_n \rangle = \langle J_n^{(\nu=1)} \rangle$ is the result of Eq.~(\ref{eq:heat_current}). $\langle J_n^{(\nu>1)} \rangle$ should then be equated for all $n$ to obtain:
\begin{equation}
\label{temp-profile}
\begin{split}
    T_{\Omega_n}^2-\frac{T_1^2 + T_2^2}{2}  
    = 
    (N+1)(\nu-1)\frac{T_{\Omega_{n+1}}^2 -2 T_{\Omega_{n}}^2 + T_{\Omega_{n-1}}^2}{4}.
\end{split}
\end{equation}
This recurrence relation can be solved (see \textit{Supplementary Material}) to find that the temperature profile is no longer flat, but for $N\gg\nu$ it exponentially converges away from the edges to the same value of $T_{\Omega_n}^2 = \frac{T_1^2 + T_2^2}{2}$ as obtained for $\nu=1$. (See $\nu=2$ calculation in Fig.~\ref{fig:heat_flow}.) In this case, we again obtain a finite (power-law suppressed) heat flow in the presence of a negligible (exponentially suppressed) temperature difference.

We note that very recently  similar results, including Eq.~(\ref{temp-profile}), have been obtained via Langevin approach in Ref.~\cite{roche2025breakdown}.

\textbf{Entropy quench at a quantum point contact:} 
Consider now the effect of reflection at a single link along the chain between islands $n-1$ and $n$: $H_{QPC} = r \cos(\varphi_n$ where $\varphi_n\equiv\varphi_n(0)=\phi_{R,n}(0)-\phi_{L,n}(0)$. We are interested here to explore the effect of this perturbation on the transport and thermodynamic properties of the array. We can use the scattering formalism 
to change the basis of the $2(N+1)$ incoming bosonic fields defined in a similar way to Eq.~(\ref{eq:incoming_vector})
such that $\varphi_n$
will be proportional to one of the new basis vectors. In this basis, the QPC will be described by a cosine perturbation acting on a single boson mode, which is known as the boundary sine-Gordon model, from which we will borrow exact results on the entropy~\cite{fendley1994exact}.

We first find the scaling dimension of $H_{QPC} = r \cos(\varphi_n)$  using standard methods~\cite{von1998bosonization}.
For a single boson mode $\phi$ ($\phi=\phi_L$ or $\phi_R$) the  scaling dimension of the operator $e^{\pm i  \phi}$ is determined by the $T=0$  correlation function  $\langle \phi(t) \phi(t') \rangle_{T=0} \sim - \log (t-t') $. The definition of the scaling dimension $\Delta$ is given as $\langle e^{i \phi(t)}  e^{- i\phi(t')}\rangle_{T=0} \propto \frac{1}{(t-t')^{2 \Delta}}$, and one obtains $\Delta [e^{\pm i \phi}]= 1/2$ corresponding to a free fermion annihilation/creation operator. Along the same lines, $\Delta [e^{\pm i \alpha \phi}]= \alpha^2/2$.
Note that in comparison to the finite-temperature current correlation function which determined the heat current in Eq.~(\ref{eq:heat_current_fluctutation}), the scaling dimension is a result of quantum fluctuations affecting the zero-temperature correlation function.

\begin{figure}
\centering
\includegraphics[width=\columnwidth]{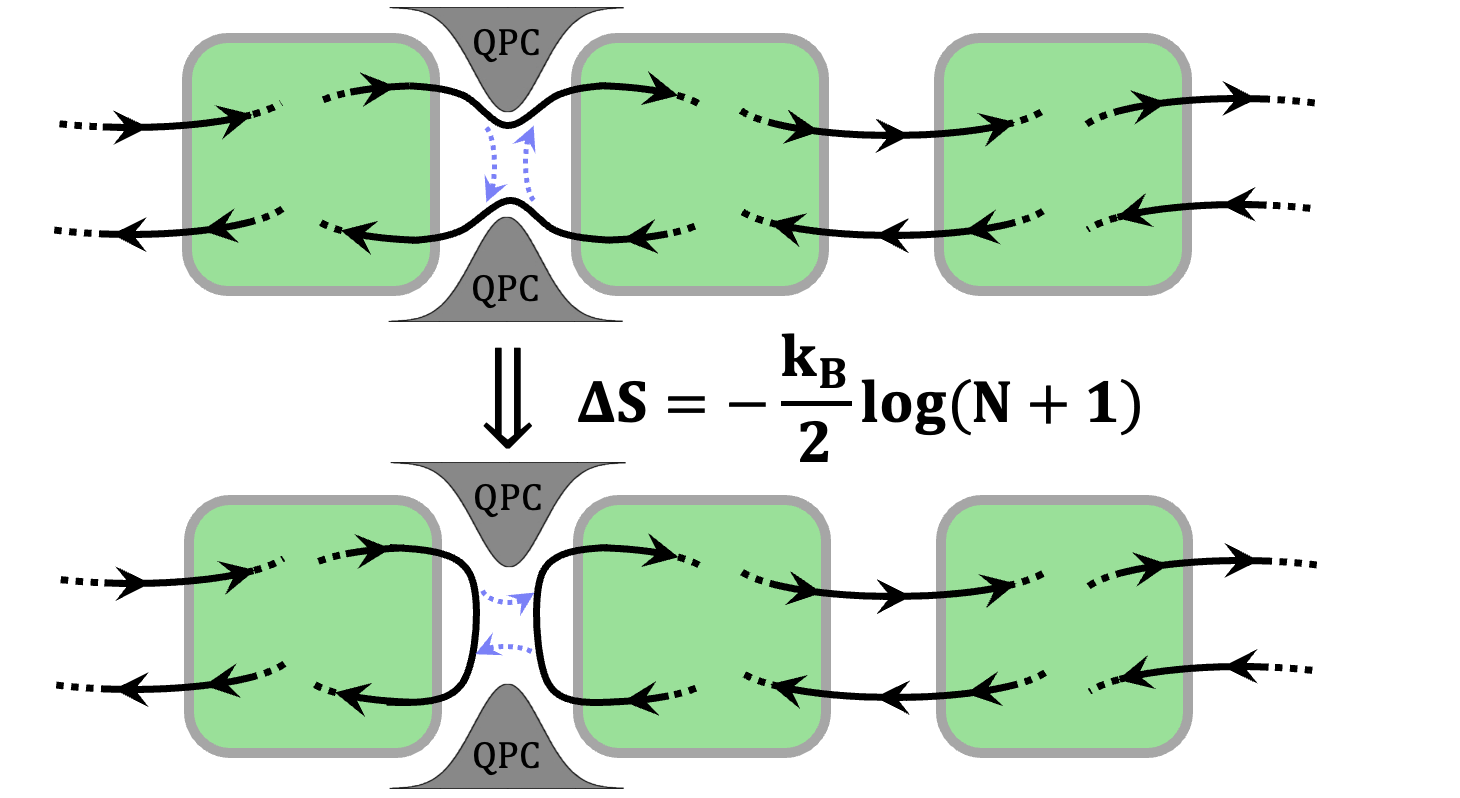}
\caption{Chain disconnection due to reflection at a QPC and associated entropy difference. }
\label{fig:1_QPC}
\end{figure}

To determine the scaling dimension of $e^{i \varphi_n}$ we need to find $\langle \varphi_n(t) \varphi_n(t') \rangle_{T=0}$. Let us first compute $\frac{e^2}{(2\pi)^2}\langle\partial_{t} \varphi_n(t) \partial_{t'} \varphi_n(t') \rangle_{T=0} =  \langle (R_n(t)-L_n(t)) (R_n(t')-L_n(t')) \rangle_{T=0} = \langle \mathcal{I}(t)\mathcal{I}(t') \rangle_{T=0} $. To find the latter current correlator, we thus want to express $\mathcal{I}$ in terms of the incoming currents, as given by Eq.~(\ref{eq:L_R}). 
Since each incoming current in Eq.~(\ref{eq:L_R}) satisfies $\langle  I_{\mu}(t) I_{\nu}(t')\rangle_{T=0} \sim - \frac{e^2}{(2\pi)^2}  \partial_{t}\partial_{t'}\delta_{\mu \nu} \log (t-t')$, we obtain 
\bea
\langle \mathcal{I}(t)\mathcal{I}(t') \rangle_{T=0}  \sim -|\vec{d}|^2  \frac{e^2}{(2\pi)^2}   \partial_{t}\partial_{t'}  \log (t-t'). \nonumber 
\eea
From here follows, 
\begin{equation}
\label{eq:Delta}
    \Delta[e^{i\varphi_n}] 
    =\frac{1}{2}|\vec{d}|^2 = \frac{1}{N+1},
\end{equation}
see \textit{Supplementary Material} for further details. 
Thus, introducing a QPC acts as a relevant perturbation with scaling dimension $\Delta = \frac{1}{N+1}$, effectively cutting the chain into two segments. Exactly as in boundary sine-Gordon model, this happens below an energy scale of $T^*_r \propto r^{(N+1)/N}$. Remarkably, due to the conservation of net current $\mathcal{I}$ through the array, and its relation to the tunneling operator, the location of the QPC does not affect this result -- the QPC can be in any of the $N+1$ links.

We now borrow an important result from the boundary sine-Gordon model: as the cosine perturbation flows from small to large coupling, there appears a universal entropy drops given by~\cite{fendley1994exact}
\be
\Delta S = \frac{1}{2} k_B \log(N+1).
\ee
Physically this entropy drop occurs as $T$ becomes smaller than the scale $T^*_r$ and the QPC goes from full transmission to full reflection, see Fig.~\ref{fig:1_QPC}. The logarithmic scaling of the entropy change with the number of islands indicates a nontrivial many-body state formed at the interacting 1D array. 

It can be shown in similar ways that for temperature $T \ll T^*_r$ that the scaling dimension of the dual operator (tunneling left-and-right instead of up-and-down) is $N+1$. This perturbation is irrelevant, meaning that the disconnected state is stable. This observation implies that the RG flow of the reflection coefficient is unidirectional, without intermediate fixed points at $0<r<1$.

Our result generalizes well-known limits. For $N = 1$, we recover the two-channel Kondo case $\Delta S = \frac{1}{2} k_B \log 2$. The $N = 2$ case, studied in Refs.~\cite{pouse2023quantum}, gives $\Delta S = \frac{1}{2} k_B \log 3$, linked to emergent $Z_3$ parafermions~\cite{karki2023z}. It is suggestive to  speculate that our result $\Delta S = \frac{1}{2} k_B \log(N+1)$ corresponds to $Z_{N+1}$ parafermions.

\begin{figure}
\centering
\includegraphics[width=\columnwidth]{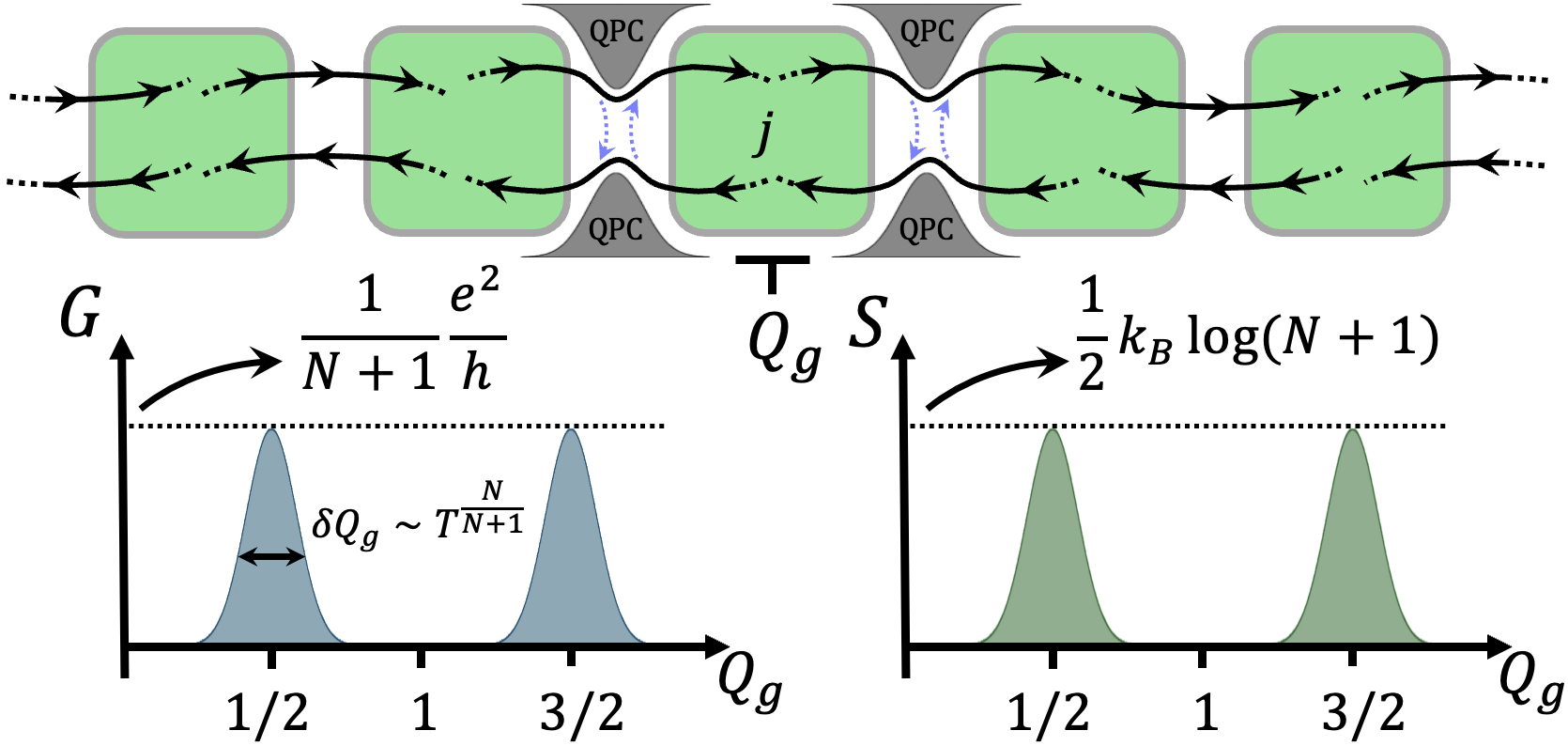}
\caption{1D array of metallic islands with 2 QPCs and schematic conductance and entropy curves. The conductance displays interference versus the gate voltage, and the entropy shows peaks scaling logarithmically with the number of islands.
}
\label{fig:2_QPCs}
\end{figure}

\textbf{Two QPCs - Interference:} 
Consider now adding a second QPC, as in Fig.~\ref{fig:2_QPCs}. The positions of the two QPCs are at links $i_1$ and $i_2$ (without the loss of generality, $1\le i_1<i_2 \le N+1$). The tunneling Hamiltonian is now $H_{\text{QPC}} = r_1 \cos \varphi_{i_1}(0) + r_2 \cos \varphi_{i_2}(0)$. 

First, let us consider this problem qualitatively. If we turn on $r_2$ at the energy scale $T \ll T^*_{r_1}$, then our starting point is two disconnected chains with $i_{1}-1$ and $N
-i_{1}-1$ islands, respectively. In this case, we find that the scaling dimension of the second, weaker QPC is $N$, which means that it is an irrelevant perturbation that scales to zero at low temperature. Consider now the more general case where the two QPCs  have similar strengths $r_1 \cong r_2$. The crucial observation is that the charging constraint enforces constant current $\mathcal{I} = R_n - L_n $ for all $n$, namely $\varphi_{i_2}(0) = \varphi_{i_1}(0)$, meaning that the two QPCs, no matter how remote they are, correspond to the exactly same operator and their coefficients simply add up. 

In order to enable a more interesting interference, we now allow for a tunable gate voltage $Q_{g,j} \equiv Q_{gate}$ applied to one of the islands located between QPC1 and QPC2, $i_1 \le j< i_2$. From the form of $H_{\rm{charging}}$ it then follows that the charging constraint becomes
\be
\varphi_{i_2}(0) = \varphi_{i_1}(0) + 2\pi Q_{gate}.
\ee
Substituting this into the Hamiltonian gives 
$H_{\text{QPC}} = \left( r_1 + r_2 e^{2\pi i Q_{gate}} \right) e^{i \varphi_{i_1}(0)} + \text{h.c.} = r_{\text{eff}} \cos \varphi(0)$.
As a result, the conductance exhibits oscillations as a function of the  gate voltages $Q_{gate}$, with maxima at half-integer values of $Q_{gate}$, see Fig.~\ref{fig:2_QPCs}. When $r_1 = r_2$, the conductance reaches $G_N$ of Eq.~(\ref{eq:G}). The width of the peak is determined by setting $T^*_{r_{\text{eff}}}=T$, where $T^*_{r_{\text{eff}}} \propto r_{\text{eff}}^{(N+1)/N}$. The exact curve can be derived using the integrability of the boundary sine-Gordon model.

As we change the gate voltage, we obtain a ``healed'' chain at half-integer values of $Q_{gate}$, and a disconnected one for the rest of the $Q_{gate}$ range. This leads to a $Q_{gate}-$dependent entropy curve, which can be measured via charge detection using the methods of Refs.~\cite{PhysRevLett.129.227702,han2022fractional,PhysRevLett.131.016601}.
Specifically, attaching a charge detector that measures $\langle Q_j\rangle$, the entropy difference between the two values of $Q_{gate}$ is given by~\cite{han2022fractional}
\begin{equation}
    \Delta S_{Q_{gate}^{(1)}\to Q_{gate}^{(2)}} = 2E_c\int_{Q_{gate}^{(1)}}^{Q_{gate}^{(2)}}  dQ_{gate} \frac{d\langle Q_j\rangle}{dT}.
\end{equation}

\textit{Conclusion.--}  In conclusion, we have derived several unexpected results of the charge fractionalization in the 1D arrays of metallic islands. These include the heat flow in the absence of thermal gradient, and fractional entropy suggestive of $Z_N$ parafermions.


\begin{acknowledgments}
\noindent{\textit{Acknowledgments.--}} ES and NH gratefully acknowledge support from the European Research Council (ERC) under the European Union Horizon 2020 research and innovation programme under grant agreement No. 951541. GF was supported by the 
U.S. Department of Energy Award No. DE-SC0002765 (thermal properties) and NSF Award No. DMR-2428579 (multi-island arrays). 
\end{acknowledgments}


\bibliography{bibliography}


\section{Supplementary Material}
\subsection{Scattering matrix}
To construct the scattering matrix Eq.~(\ref{eq:S}), we define a basis of incoming fields $\tilde{I}_i$, being either a charge mode or a neutral mode,
\begin{equation}
    \label{eq:scatter}
    \vec{\tilde{I}}
    = \frac{1}{\sqrt{2}}\begin{pmatrix}
    I_1 + I_2 \\
    I_1 - I_2 \\
    I_3 + I_4 \\
    I_3 - I_4 \\
    \end{pmatrix}
    = \frac{1}{\sqrt{2}}\begin{pmatrix}
    1 & 1 & 0 & 0\\
    1 & -1 & 0 & 0\\
    0 & 0 & 1 & 1\\
    0 & 0 & 1 & -1\\
    \end{pmatrix}
    \begin{pmatrix}
    I_1 \\
    I_2 \\
    I_3 \\
    I_4 \\
    \end{pmatrix}
    \equiv U \vec{I},
\end{equation}
and similarly for the outgoing fields $\vec{\tilde{O}}=U \vec{O}$. Notice that $U$ is unitary and symmetric.
Following Morel et. al.~\cite{morel2022fractionalization} we state that '+' superpositions (charged) are fully reflected, and '-' superpositions (neutral) are fully transmitted. Therefore, in this basis, the scattering relation $\vec{\tilde{O}} = \tilde{S}\vec{\tilde{I}}$ is given by
\begin{equation}
    \tilde{S} = \begin{pmatrix}
    0 & 0 & 1 & 0\\
    0 & 1 & 0 & 0\\
    1 & 0 & 0 & 0\\
    0 & 0 & 0 & 1\\
    \end{pmatrix}.
\end{equation}
This can be decomposed into an identity in the space of '+' superpositions and a  $\sigma_x$ matrix in the space of '-' superpositions.

The scattering matrix in the original basis is thus 
\begin{equation}
    S = U\tilde{S}U 
    = \frac{1}{2}\begin{pmatrix}
    1 & -1 & 1 & 1\\
    -1 & 1 & 1 & 1\\
    1 & 1 & 1 & -1\\
    1 & 1 & -1 & 1\\
    \end{pmatrix},
\end{equation}
which is Eq.~(\ref{eq:S}).


\subsection{Derivation of the current recurrence relation, Eq.~(\ref{eq:sol_recurrence})}
Expressing the currents through any channel in terms of independent incoming fields, as obtained in Eq.~(\ref{eq:sol_recurrence}), is a key result in the main text. Here we  derive it explicitly.

Going back to Eq.~(\ref{eq:recurrence}) we can solve it by induction for the ranges $1\rightarrow (n-1)$ and $n\rightarrow (N+1)$ to find the  equations
\begin{subequations}
\begin{equation}
    L_1 = L_n + (n-1)\mathcal{I} - \sum_{i=1}^{n-1}I^{(isl)}_i,
\end{equation}
\begin{equation}
    L_n = L_{N+1} + (n'-1)\mathcal{I} - \sum_{i=n}^{N}I^{(isl)}_i,
\end{equation}
\end{subequations}
with $n' = (N+1)-n+1$ being the index $n$ counted from the end of the array. Then using $\mathcal{I} = R_n-L_n$ (and in particular $R_n-L_n = R_1-L_1$) we can change those equations to the symmetric form of
\begin{subequations}
\begin{equation}
    R_n = \frac{1}{n}((n-1)L_n + R_1 + \sum_{i=1}^{n-1}I^{(isl)}_i),
\end{equation}
\begin{equation}    
    L_n = \frac{1}{n'}((n'-1)R_n + L_{N+1} - \sum_{i=m}^{N} I^{(isl)}_i),
\end{equation}
\end{subequations}
which is a pair of linear equations with two variables $R_n, L_n$. After some careful algebra the solution, i.e. Eq.~(\ref{eq:sol_recurrence}), is reached.


\subsection{Heat current and temperature profile - Two islands}
In the main text we presented the method of calculating the heat current and temperature profile, and displayed it explicitly for the case of a single island. We now proceed to the two island case.
The currents that are not incoming currents - $L_1, R_3, L_2, R_2$ - are given by Eq.~(\ref{eq:sol_recurrence}) so the heat currents are then immediately derived by $\langle J_n \rangle  = \langle J_{R_n} \rangle - \langle J_{L_n} \rangle$ using Eq.~(\ref{eq:heat_current_fluctutation}),
\begin{equation}\begin{split}
    L_1 = \frac{1}{3}(L_3 &+ 2R_1 - I^{(isl)}_1 - I^{(isl)}_2) 
    \\~~\rightarrow~~
    \langle J_{L_1} \rangle 
    &= \frac{\kappa}{9}(T_2^2 + 4T_1^2 + 2T_{\Omega_1}^2+ 2T_{\Omega_2}^2).
\end{split} \end{equation}

Since $R_1$ is an incoming current from the left reservoir, it transfers a heat current of $\langle J_{R_1} \rangle = \kappa T_1 ^2 $. Then the heat current through link 1 is
\begin{equation}\begin{split}
    \langle J_1 \rangle = \langle R_1^2 \rangle - \langle L_1^2 \rangle
    = \frac{\kappa}{9}(5T_1^2 - T_2^2 - 2T_{\Omega_1}^2 - 2T_{\Omega_2}^2).
\end{split} \end{equation}
Using the left-right symmetry we can find the same heat current from the other side,
\begin{equation}\begin{split}
    \langle J_3 \rangle 
    = \frac{1}{9}(T_1^2 - 5T_2^2 + 2T_{\Omega_1}^2+ 2T_{\Omega_2}^2).
\end{split} \end{equation}
The last heat current $\langle J_2 \rangle $ is derived similarly (using Eq.~(\ref{eq:sol_recurrence}) for both $R_2$ and $L_2$), giving
\begin{equation}
    \langle J_2 \rangle 
    = \frac{1}{3}(T_1^2 - T_2^2 + 2T_{\Omega_1}^2 - 2T_{\Omega_2}^2).
\end{equation}
The conservation of heat flow $\langle J_1 \rangle = \langle J_3 \rangle$ leads to $T_1^2 + T_2^2 = T_{\Omega_1}^2 + T_{\Omega_2}^2$. Plugging this into  $\langle J_1 \rangle = \langle J_2 \rangle$ we get the constant temperature profile,
\begin{equation}
T_{\Omega_1} = T_{\Omega_2} = \frac{T_1^2 + T_2^2}{2}.    
\end{equation}
Substituting back into one of the $\langle J_i \rangle$ we find that the heat current is $J = \frac{T_1^2 - T_2^2}{3}$.


\subsection{Solution to the temperature recurrence relation (Eq.~(\ref{temp-profile}))}
We notice first that Eq.~(\ref{temp-profile}) is a discrete Poisson equation. Thus the factor of $\frac{(N+1)(\nu-1)}{4}$ is analogous to an inverse squared "length" $\ell$, such that $\ell = 2/\sqrt{(N+1)(\nu-1)}$. By defining a relative squared temperature $t_n = T_n^2-\frac{T_1^2 +T_2^2}{2}$ we go from a Poisson form to a standard recurrence relation form
\begin{equation}\begin{split}
    \frac{1}{\ell^2}(t_{n+1} &- 2t_n + t_{n-1})  = t_n
    \\ &\to t_{n+1} - t_n (2 + \ell^2) + t_{n-1} = 0.
\end{split}\end{equation}
Using the ansatz $r^n$ we find the characteristic equation $r^2 - (2+\ell^2)r + 1 = 0$ with solutions  
\begin{equation}
    r_\pm = 1 + \frac{\ell^2}{2} \pm \sqrt{\left(\frac{\ell^2}{2}\right)^2  +\ell^2}.
\end{equation}
As long as $\ell\neq0$ i.e. $\nu\neq1$, the solution is of the form
\begin{equation}\begin{split}
    t_n = A_+ r_+^n + A_- r_-^n,
\end{split}\end{equation}
with coefficients $A_+$ and $A_-$  that depend on the boundary conditions. Notice that $r_+>1$ and $r_-<1$.
Imposing symmetric boundary conditions such that $t_0 = 0$ and $t_{-N/2} = t_1 = - t_2$, the result is
\begin{equation}
    t_n = \frac{t_1}{r_+^{-N/2} - r_-^{-N/2}} (r_+^n - r_-^n),
\end{equation}

\begin{equation}
    t_n = \frac{t_1}{r_+^{-N/2} - r_-^{-N/2}} (r_+^n - r_-^n).
\end{equation}
Returning to indices $1\to N$ and to the original temperatures we reach the exact solution
\begin{equation}
\label{eq:exact_T}
    T^2_n = \frac{T_1^2 +T_2^2}{2} + \frac{(T_1^2 - T_2^2)/2}{r_+^{-N/2} - r_-^{-N/2}} (r_+^{m-N/2} - r_-^{m-N/2}).
\end{equation}
Fig.~\ref{fig:heat_flow} shows graphical results for the above equation for $N=1000,~~\nu=2,100$.\\

\subsection{Another derivation of the scaling dimension Eq.~(\ref{eq:Delta})}

\begin{figure*}[t]
\centering
\includegraphics[width=0.6\textwidth]{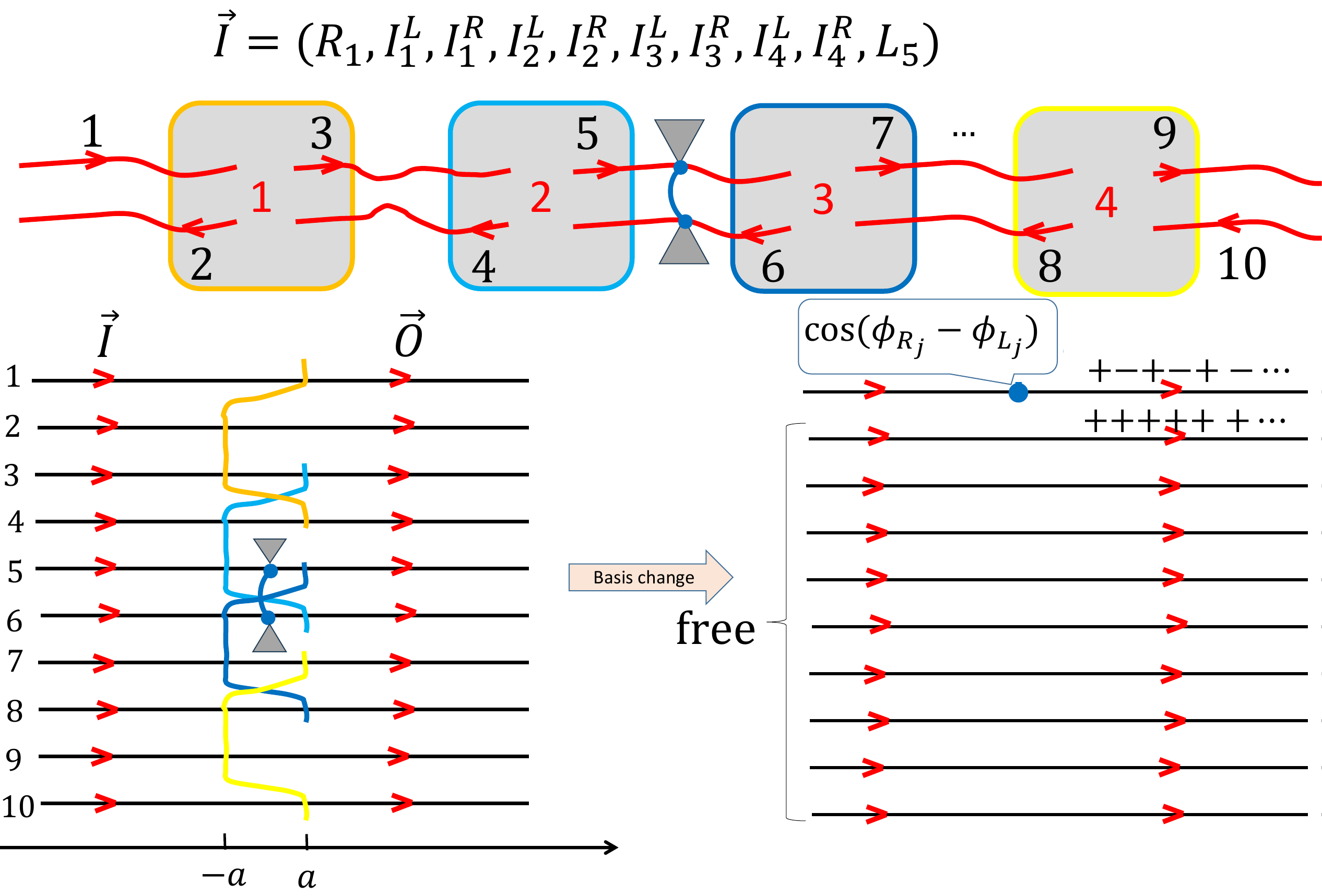}
\caption{(a) We illustrate explicitly the 10 components of the incoming current vector $\vec{I} = \{I_\mu \}$ for the case of 4 islands. (b) Rather than talking about currents $j_\mu = -\frac{e}{2\pi} \partial_t \phi_{\mu}$, one can equivalently distinguish incoming and outgoing components of the bosonic fields $\{ \phi_{\mu} \}$. We  draw all the boson fields as right movers propagating from the incoming- to the outgoing-region, and depict the interactions due to the separate charging energy terms. The QPC acts at an intermediate stage within the scattering process. (c) The particular combination of incoming fields $\tilde{\phi}_1 = \frac{1}{\sqrt{2(N+1)}}\sum_{\mu=1}^{2(N+1)} (-1)^\mu \phi_\mu$, corresponding to Eq.~(\ref{eq:L_R}), is a globally neutral scattering eigenmode to which the QPC uniquely couples at any link.
} 
\label{fig:basis_change}
\end{figure*}

In deriving Eq.~(\ref{eq:Delta}) we referred to the fact that using the scattering formalism one can change basis in the $2(N+1)$-dimensional space of boson fields such that the QPC perturbation acts on a single decoupled component. This is illustrated in Fig.~\ref{fig:basis_change} for $N=4$. 

Consider the orthonormal combination of incoming fields (rather than currents),
\be
\tilde{\phi}_1 = \frac{1}{\sqrt{2(N+1)}}\sum_{\mu=1}^{2(N+1)} (-1)^\mu \phi_\mu.
\ee
Due to the equality in Eq.~(\ref{eq:L_R}), the QPC operator at any link coupled exactly to this field. However there is a coefficient, $\cos( \varphi_n) = \cos (\alpha \tilde{\phi}_1)$ where $\alpha =\sqrt{\frac{2}{N+1}}$. The scaling dimension is then $\Delta = \frac{1}{2}\alpha^2 = \frac{1}{N+1}$.

\end{document}